\documentclass[aps,pre,onecolumn,10pt,floatfix,superscriptaddress,preprint]{revtex4-2}

\usepackage{sidecap}
\usepackage{afterpage}
\usepackage{ragged2e}
\usepackage{graphicx}
\usepackage{amsmath, bm, mathtools}
\usepackage{physics}
\usepackage{amssymb}
\usepackage{xcolor}
\usepackage[utf8]{inputenc}
\usepackage{tikz}
\usepackage{braket}
\usepackage[normalem]{ulem}
\usepackage{lineno} 
\usepackage{hyperref}

\usepackage{setspace}

\usepackage[]{natbib}

\PassOptionsToPackage{linktocpage}{hyperref}

\graphicspath{{./Figures}{}}

\usepackage{xcolor}

\newcommand{\quotes}[1]{``#1''}

\begin{document}


\title{Component systems: do null models explain everything?} 





\author{Andrea Mazzolini}
\email{andrea.mazzolini.90@gmail.com}
\affiliation{Dipartimento di Fisica, Universit\`a di Torino, Turin, Italy}
\affiliation{I.N.F.N, Turin, Italy}
\affiliation{I.N.F.N, Bari, Italy}

\author{Mattia Corigliano}
\affiliation{IFOM ETS, the AIRC Institute of Molecular Oncology,  Milan, Italy}

\author{Rossana Droghetti}
\affiliation{IFOM ETS, the AIRC Institute of Molecular Oncology,  Milan, Italy}

\author{Matteo Osella}
\affiliation{Dipartimento di Fisica, Universit\`a di Torino, Turin, Italy}
\affiliation{I.N.F.N, Turin, Italy}

\author{Marco Cosentino-Lagomarsino}
\email{marco.cosentino-lagomarsino@ifom.eu}
\affiliation{IFOM ETS, the AIRC Institute of Molecular Oncology,  Milan, Italy}
\affiliation{Dipartimento di Fisica, Universit\`a degli Studi di Milano, Milan, Italy}
\affiliation{I.N.F.N, Milan, Italy}

\date{\today}

\begin{abstract}
Component systems --- ensembles of realizations built from a shared repertoire of modular parts --- are ubiquitous in biological, ecological, technological, and socio-cultural domains. 
From genomes to texts, cities, and software, these systems exhibit statistical regularities that often meet the \emph{bona fide} requirements of laws in the physical sciences. 
Here, we argue that the generality and simplicity of those laws are often due to basic combinatorial or sampling constraints, raising the question of whether such patterns are actually revealing system-specific mechanisms   and how we might move beyond them. 
To this end, we first present a unifying mathematical framework, which allows us to compare modular systems in different fields and highlights the common ``null'' trends as well as the system-specific uniqueness, which, arguably, are signatures of the underlying generative dynamics. 
Next, we can exploit the framework with statistical mechanics and modern machine-learning tools for a twofold objective. (i) Explaining why the general regularities emerge, highlighting the constraints between them and the general principles at their origins,  and (ii) ``subtracting`` them from data, which will isolate the informative features for inferring hidden system-specific generative processes, mechanistic and causal aspects.
\end{abstract}

\maketitle

\section*{Introduction}

The study of complex systems has emerged as a central theme in modern
science, driven by the recognition that many phenomena, from
biological systems to technological infrastructures, share underlying
organizational principles. A particularly productive framework for
understanding these systems is the concept of modularity. Modularity
refers to the decomposition of a complex system into smaller,
relatively independent components that interact to produce
system-level behavior~\cite{newman2006modularity}. This perspective on empirical data is not new;
it has been applied in fields ranging from biology~\cite{Koonin2002,Lapenta2020} to
software engineering~\cite{Koch2007} and even urban
planning~\cite{barthelemy2011spatial}. In the last decades, advances in data
availability, computational power, and theoretical tools have opened
new avenues for exploring modularity in unprecedented detail, as well as for understanding its origins~\cite{Kashtan2005,Wagner2007,Lorenz2011}.

Across domains as diverse as evolutionary genomics,  linguistics,
software engineering, and urban studies, complex systems can often be
represented as realizations assembled from modular components, or
``component systems''. Genomes consist of gene families; texts consist
of words; microbiomes are assemblies of individuals from different 
taxa; LEGO kits consist of plastic bricks; and single-cell
transcriptomes consist of expressed genes. Despite their
heterogeneity, these systems reveal common empirical regularities.
Well-known examples include Zipf's law for component frequencies~\cite{zipf1949human,li2002zipf,corominas2010universality,lu2010zipf,baek2011zipf,aitchison2016zipf,mazzolini2018statistics,piantadosi2014zipf}, the  
sublinear power-law-like scaling of diversity -or vocabulary size-  with system size (known
as Heaps' law in linguistics)~\cite{heaps1978information,cosentino2009universal,mazzolini2018statistics,mazzolini2018heaps}, and the widespread U-shaped occurrence  patterns of gene families in genomics~\cite{koonin2008genomics,touchon2009organised,mazzolini2018statistics}. 

These patterns have been discovered independently in several fields, and system-specific generative processes have been proposed to explain their emergence. The common mathematical framework proposed in this Review can help the recognition of the generality of these patterns and promote the fruitful exchange of ideas and data analysis techniques between fields.

\begin{figure}[htpb]
    \centering
    \includegraphics[width=0.95\textwidth]{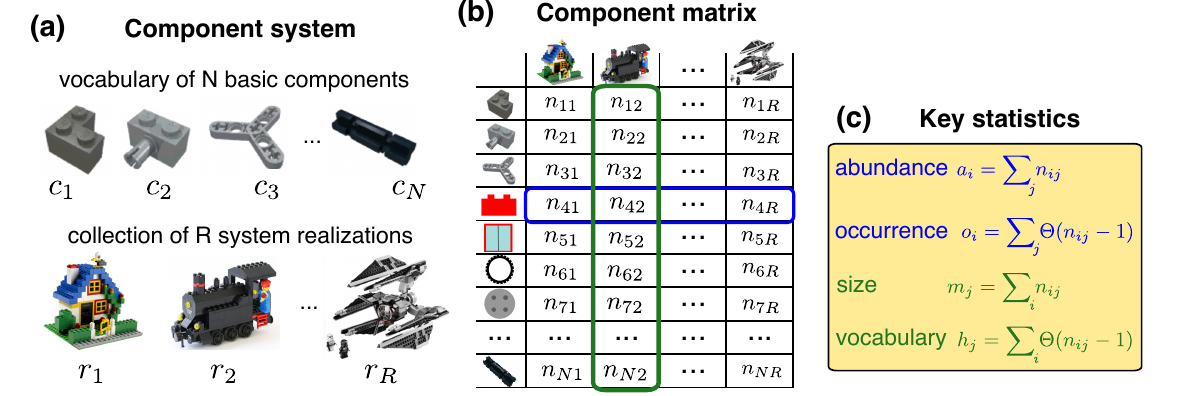}
    \caption{
    \textbf{Component systems as a unifying representation of modular artifacts.}
    (a) A wide class of systems---including genomes, texts, and LEGO constructions---can be described as \emph{component systems}, in which each realization is assembled from a shared vocabulary of elementary components that may be reused within and across realizations.
    (b) This structure is encoded by the component matrix $n_{ij}$, whose entries count how many times component $i$ appears in realization $j$.
    (c) Simple sums and binarized sums of the matrix elements define a set of fundamental observables that characterize the system at both the component and realization level, including component abundance and occurrence, as well as realization size and vocabulary.
    }   
    \label{fig:fig1}
\end{figure}

\section*{The component system and its emerging laws}

A unifying description of component systems reveals that their apparent complexity is governed by a small set of robust and reproducible statistical regularities. These patterns, often described as emerging ``laws'', recur across systems~\cite{van2003scaling,Koonin2011,cosentino2009universal,grilli2012joint,mazzolini2018statistics,lazzardi2023emergent,grilli2020macroecological,youn2016scaling} with widely differing microscopic details. Before entering the details of these results, we first introduce a common notation.

\subsection*{A unifying notation for component systems}

The basic feature of a modular or component system is that its  \textit{realizations} --- genomes, books, LEGO sets --- are made of elementary \textit{components} --- genes, words, LEGO bricks --- that can be reused within and between realizations~(Fig.~\ref{fig:fig1}(a)).
This can be encoded in a matrix of integer elements $n_{ij}$, indicating how many times the component $i$ appears in the realization $j$~(Fig.~\ref{fig:fig1}(b)).

Starting from this matrix, we introduce the first set of definitions involving summations of rows or columns~(Fig.~\ref{fig:fig1}(c)).
The sum within a realization, $m_j = \sum_i n_{ij}$, defines its \textit{size} in terms of how many components it contains, and it can be used to normalize the counts defining the frequency $f_{ij} = n_{ij} / m_{j}$ (Table~\ref{tab:math_symbols}).
The sum across realizations of a given component, $a_i = \sum_j n_{ij}$, is the \textit{abundance} of that component in the ensemble.
The component abundances are often normalized to obtain an ensemble frequency $g_i = a_i / \sum_{i'} a_{i'}$.

A next set of informative quantities (Table~\ref{tab:math_symbols}) involves the binarized matrix that describes the presence-absence patterns. First we have the \textit{sharing number} or \textit{occurrence} of a component that counts in how many realizations the component is present, $o_i = \sum_j \Theta(n_{ij}-1)$, where $\Theta(x)$ is the Heaviside function, which is $1$ for $x > 1$.
Its transposed quantity is the \textit{vocabulary} of a component, indicating the number of different components in the realization, $h_j = \sum_i \Theta(n_{ij}-1)$.

At this level of description, the component system is a weighted bipartite network \cite{diestel2025graph} with an adjacency matrix $n_{ij}$,  where abundance and size correspond to the node degrees on the two layers.
This parallel allows us to benefit from many tools that have been introduced in network theory. 
For example, measures of node centrality that define the \quotes{importance} of a component or realization~\cite{tacchella2012new, mazzolini2025ranking}, or community detection and topic modeling techniques to find groups of realizations characterized by similar component compositions~\cite{blei2003latent, gerlach2018network, valle2020topic,valle2025exploring,valle2022multiomics}.
However, we argue that a component system deserves a specific notation for two main reasons.
The first is conceptual: the class of systems that we are describing here have a precise interpretation of the two layers --- realizations and components --- and meaning of the introduced quantities --- abundance, size, sharing number and vocabulary ---, that would be lost with the network notation.
The second reason is the fact that a component system has an additional set of properties described below that go beyond bipartite networks.

\begin{table*}
	\centering
	\begin{tabular}{c|c} 
		$n_{ij}$ & count of component $i$ in realization $j$ \\ \hline
		$f_{ij}$ & frequency of component $i$ in realization $j$ \\ \hline
        $m_{j}$ & size of realization $j$ \\ \hline
        $h_{j}$ & vocabulary of realization $j$ \\ \hline
		$a_{i}$ & abundance of component $i$ \\ \hline
		$g_{i}$ & ensemble frequency of component $i$ \\ \hline
		$o_{i}$ & occurrence/sharing-number of component $i$ \\
	\end{tabular}
	\caption{List of main mathematical symbols of a component system}
	\label{tab:math_symbols}
\end{table*}

Components can often be grouped into classes that play similar roles in determining the functioning of a realization. For instance, protein-coding genes may be classified according to structural or evolutionary criteria, or into  coarser functional families, while words can be grouped into syntactic or semantic categories. This hierarchical organization enables the definition of macro-components and allows the same system to be described at multiple levels of resolution. Such multiscale structure is a defining feature of component systems and introduces additional constraints and observables that are not naturally captured by standard bipartite network representations.
At the same time, realizations can have a natural hierarchical structure, as genomes can be grouped into taxonomical units, books into different topics and LEGOs into different themes.
As discussed later, the study of how the system statistics change by coarse-graining at the level of components and realizations is still under-explored.

Closely related to this multiscale organization is the presence of dependencies among components, which constrain their joint occurrence within realizations~\cite{mazzolini2018zipf}. In many systems, a given module can appear only if other modules are present, as exemplified by software dependencies in Linux distributions~\cite{pang2013universal}. More generally, one can posit the existence of an (often hidden) dependency network among components, whose non-trivial structure gives rise to functional clusters that naturally correspond to the macro-components introduced above.

A final additional structure, present only in some component systems, is the temporal ordering of components. This ordering is naturally defined in linguistic artifacts, where words follow a well-defined sequence, while it is more blurred or even inaccessible in other systems. For instance, in LEGO constructions the order of bricks may be associated with the sequence of assembly steps, yet multiple construction paths can lead to the same final composition, and such information is rarely available in public datasets. The challenge is even more pronounced in genomic systems, where ordering may reflect evolutionary times of gene acquisition, but its precise definition and empirical reconstruction remain elusive. Nevertheless, when a meaningful notion of order can be identified, it enables the definition of additional observables and statistics that encode valuable information about the system’s organization and generative mechanisms~\cite{altmann2009beyond,altmann2012origin,mazzolini2018heaps}.

\begin{figure}[htpb]
    \centering
    \includegraphics[width=0.95\textwidth]{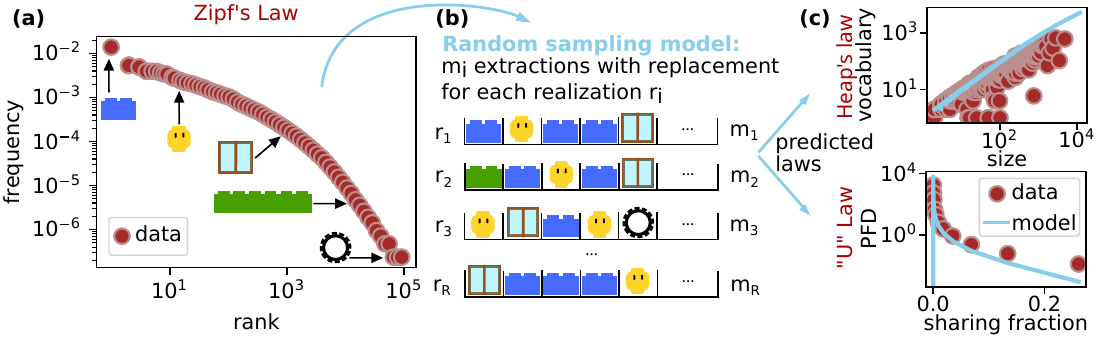}
    \caption{
    \textbf{Basic statistical laws and their connection due to sampling constraints.  }
    (a) Zipf's law for component frequency is reported for the LEGO dataset (\url{https://rebrickable.com/}).
    (b) By extracting components with their empirical probabilities, a random sampling model can generate an artificial ensemble of realizations that can be compared to the empirical ones.  
    (c) The statistics of shared components and the Heaps' law can be often explained by this sampling procedure, although quantitative deviations can reveal system-specific mechanisms. 
    }   
    \label{fig:fig2}
\end{figure}

\subsection*{The emerging \quotes{physical laws} of component systems}

The definition of a common language between different systems opens up the possibility of exporting concepts and results between fields.
In particular, the study of emerging patterns has a long tradition in linguistics \cite{altmann2016statistical} and ecology \cite{brown1995macroecology, grilli2020macroecological},   and  can now be systematically extended to many other systems.
The fascination of these patterns or laws is due to their mathematical simplicity and robustness, while the underlying systems are complex and heterogeneous.
In this regard, extending the range of systems in which those regularities can be comparatively studied provides  more hints about general and robust features or, instead, system-specific behaviors~\cite{mazzolini2018statistics,lazzardi2023emergent}.
Before going to the critical discussion about the information these \quotes{laws} carry and how to exploit them to better understand a system, below we present a list of the more studied patterns.

\subsubsection*{Zipf, Heaps, and component-sharing laws}

A first class of empirical regularities concerns statistics obtained from the row and column sums of the component matrix $n_{ij}$, as well as of its binarized counterpart (see Tab.~\ref{tab:math_symbols}). Among these patterns, the most extensively documented is Zipf's law, originally identified in linguistics~\cite{zipf1949human} and later observed across a wide range of systems~\cite{li2002zipf, newman2005power, lazzardi2023emergent, mazzarisi2021maximal}. Zipf's law describes the distribution of component frequencies, either at the ensemble level through $g_i$ or, depending on the context, within individual realizations through $f_{ij}$. 

Defining the rank $r_i$ of a component as its position in the list of components ordered by decreasing frequency (with $r_i = 1$ denoting the most frequent component), Zipf's law states that $g_i \propto r_i^{-\mu}$, where the exponent $\mu$ is typically close to unity (Figure~\ref{fig:fig2}(a)). The extensive literature devoted to this regularity, as well as to the various mechanisms proposed to explain it, is reviewed elsewhere~\cite{mitzenmacher2004brief, newman2005power, piantadosi2014zipf}.

Shifting the focus from component-level to realization-level statistics naturally leads to Heaps' law, originally introduced in the context of quantitative linguistics~\cite{herdan1964quantitative, heaps1978information, egghe2007untangling}, and subsequently extended to other domains, including genomics~\cite{cosentino2009universal} and innovation dynamics~\cite{tria2014dynamics, loreto2016dynamics}. Heaps' law characterizes how the vocabulary size $h_j$ of a realization grows with its total size $m_j$, capturing the progressive slowdown in the introduction of new components as realizations become larger. Empirically, the average vocabulary size at a given realization size follows a sublinear scaling, $\langle h \rangle \propto m^\nu$, with $0 < \nu < 1$ (Figure~\ref{fig:fig2}(c)). 
Recent studies have also addressed the fluctuations around this average scaling behavior~\cite{gerlach2014scaling, tria2020taylor, lazzardi2023emergent}, revealing a robust quadratic scaling of the variance, $\sigma_h^2 \propto \langle h \rangle^2$, in agreement with Taylor's law~\cite{taylor1961aggregation}.

A third class of regularities concerns the sharing statistics of components, quantified by the occurrence \(o_i\), defined as the fraction of realizations in which a given component is present. The distribution of \(o_i\) has been extensively studied in comparative genomics~\cite{touchon2009organised, koonin2008genomics, lobkovsky2013gene}, where it exhibits a characteristic U-shaped form. This structure reflects the coexistence of a small set of highly conserved ``core'' components, present in nearly all realizations (\(o_i \simeq 1\)), alongside a large number of ``accessory'' or rare components that appear in only a small fraction of realizations (\(o_i \sim 1/R\), where \(R\) denotes the total number of realizations).
A related example arises in quantitative immunology, where the sharing of T-cell receptor clones across individuals is analyzed. In this context, the occurrence distribution has been used to test and validate generative models of receptor sequences, whose generation probabilities are strongly heterogeneous.
This result holds in \quotes{public} receptors, which are easier to produce and likely shared between many individuals, as opposed to \quotes{private} clones, which tend to be unique to only one person~\cite{elhanati2018predicting, ruiz2023modeling}.

\subsubsection*{Beyond the statistics of row and column summations}

Statistics that go beyond simple row and column summations of the component matrix typically lack the universality and robustness of the aforementioned laws, and often display substantial heterogeneity both within and across component systems. For this reason, they are comparatively less prominent in the literature, despite being potentially more informative about system-specific organization and underlying generative mechanisms.

In genomics, for example, Grilli and coworkers~\cite{grilli2014cross} analyzed the statistics of the frequencies $f_{ij}$ by fixing a given gene family $i$ and studying its distribution across genomes.
They found that the resulting distribution shapes vary widely among gene families, and linked this variability to differing evolutionary processes, including rates of horizontal gene transfer and family-specific expansion dynamics.

In linguistics, a substantial body of work has explored more refined statistical observables~\cite{altmann2016statistical}. One example, which may be relevant to other component systems, is the tendency of rare words at the ensemble level to cluster within specific documents while being absent from most others~\cite{serrano2009modeling}, a behavior that deviates from null models based on independent sampling from ensemble frequencies. Additional approaches focus on sequential properties, such as the entropy of word sequences~\cite{ebeling1994entropy}, or more generally on statistics that explicitly account for temporal order~\cite{altmann2009beyond}. A variety of statistical and probabilistic models have been proposed to reproduce these patterns and to serve as suitable null generative models for data analysis and mining, including Poisson-based and bursty-process frameworks~\cite{church1995poisson, kleinberg2002bursty}, but the debate remains open regarding the selection among different candidate scenarios.

\subsubsection*{Including functional categories}

In principle, all the discussed statistics can be studied by coarse-graining the matrix of counts through \textit{a priori} grouping of components into functional categories or by grouping realizations based on similarity.
There are examples of studies of this type in microbial ecology~\cite{sireci2023environmental}, where specific marginal statistics can be explored by varying the phylogenetic similarity between species.

In genomics, different works have studied the scaling relation between the amount of gene families belonging to a macro-category, e.g., genes involved in metabolism or transcription factors, and  the genome size~\cite{van2003scaling,Molina2009,de2017family}.
The scaling is a power-law-like growth with a category-dependent exponent, which can be understood in terms of basic assembly rules that depend on the category function~\cite{maslov2009toolbox, Molina2008,grilli2012joint}. 
Crucially, this behavior cannot be obtained by sampling models, and is not found in most other empirical component systems~\cite{mazzolini2018statistics,de2017family}. 
Interestingly, the same type of nonlinear category-specific scaling has been also reported for socio-economic units within cities, e.g., transportation, scientific units, agricultural activities and many more~\cite{youn2016scaling}, showing, again, a category dependent growth as the city size increases.
In that context, different activity classes were argued to follow sublinear, linear, or superlinear scaling, reflecting their position within an urban functional hierarchy, an interpretation that is remarkably similar to the ones proposed for the genomic counterpart~\cite{maslov2009toolbox, grilli2012joint}.
For urban data, the observed saturation of diversity in the largest cities was interpreted as finite classification resolution, rather than from an intrinsic limit to diversification, pointing to an underlying open-ended growth of functional diversity with system size.

\section*{Key regularities emerge from sampling constraints}

A central outcome of recent work on component systems, ranging from genomes to software, is the identification of robust null trends that emerge even in the absence of specific biological or functional constraints. In particular, equilibrium sampling models have been shown to reproduce several empirical regularities. In these models, realizations are generated by drawing components from a fixed frequency distribution, in close analogy with ensembles of random networks with a prescribed degree sequence~\cite{park2004statistical, cimini2019statistical}. 

In its simplest formulation, the sampling probabilities are taken to be equal to the empirical Zipf's law frequencies $g_i$, and each realization is composed of a number of independently drawn components equal to its size $m_j$ (Fig.~\ref{fig:fig2}(b)). Despite its minimal assumptions, this framework provides a natural explanation for the sublinear growth observed in Heaps' law (Fig.~\ref{fig:fig2}(c)), predicting an average scaling $\langle h \rangle \propto m^{1/\mu}$, where $\mu$ denotes the Zipf exponent~\cite{van2005formal, lu2010zipf, eliazar2011growth}. 

The same null model also yields quantitative predictions for the statistics of shared components across realizations. Taken together, these results demonstrate that strong connections between different marginal observables—such as Zipf's law, Heaps' law, and component sharing—can arise purely from sampling effects. Empirically, a wide range of component systems are found to approximately follow these predictions (Fig.~\ref{fig:fig2}(c))~\cite{mazzolini2018statistics}, indicating that information is often redundant across laws. This redundancy suggests that different regularities, as well as different systems, should be analyzed jointly rather than in isolation.

The sampling model above is just an example of statistical-mechanics models that can describe the consequences of a given input hypothesis, in this case the Zipf's law for the universe of components, on the different statistical laws of the system.
Going beyond a simple sampling, one can add constraints where components carry intrinsic limitsm, cross-enrichment or self-enrichment in their abundance. Models of this kind can explain features such as universal low-abundance yet ubiquitous gene families—can arise from occupancy~\cite{leinaas1977theory, grilli2014cross}. 
Similarly, partition statistics derived from \quotes{least constrained} ensembles recapitulate macro-level scaling patterns across functional categories~\cite{van2003scaling} and provide baseline expectations for condensation-like phenomena, where a small number of components dominate system-wide usage~\cite{bianconi2001bose, bassetti2009statistical}. 
These results highlight that a surprisingly large portion of large-scale regularities emerge from a heterogeneous component abundance usage, described by the Zipf's law, in combination with sampling and combinatorial structure alone, before introducing any explicit evolutionary or functional coupling.

\section*{Subtracting regularities unveils system-specific mechanistic and causal processes}

But if these sampling models account for so many large-scale patterns across such different systems, does this make them irrelevant, or worse, trivial? And how can we then go beyond them?  
The answer lies in two complementary outcomes. 
First, null models reveal what is common and unifying across modular systems --- from genomes to software —-- by identifying the statistical signatures that arise from key general ingredients and constraints, related to combinatorics, generative rules (e.g. use of module copy-pasting \cite{qian2001protein, cosentino2009universal}), statistical dependencies, and system size. 
This allows us to identify and develop an intuition regarding the general behavior shared by diverse architectures, providing a baseline notion of universality. 
Second, once this baseline is established, we can identify and infer the system-specific mechanisms that deviate from the null expectations: dependency hierarchies, evolutionary constraints, functional specialization, correlated innovations, and ecological pressures. 
These models act as a magnifying lens, allowing principled inference of finer structure that would otherwise be obscured by the overwhelming general trends.

Against this clarified baseline, key specific model ingredients, for
example dependency structures and history-dependent novelty
processes~\cite{cosentino2009universal, tria2014dynamics, corominas2010universality, mazzolini2018zipf}
introduce deviations that match empirical signatures across
systems. For example, dependency-structure based models treat genomes
or software systems as directed acyclic graphs in which components
require other components to be functional~\cite{mazzolini2018zipf}. Imposing
these asymmetries alters both the abundance distributions and their
scaling with system size, yielding patterns that cannot be reproduced
by equilibrium sampling alone \cite{pang2013universal}. These enriched models
produce key testable predictions linking module observables to
underlying dependency hierarchies.
%

When null models are coupled with inference frameworks, deviations
themselves become informative. For genomes, deviations in family-size
scaling can indicate evolutionary potentials, dependency structures,
or horizontal gene transfer. In software, they may reveal
architectural constraints, robustness bottlenecks, or organizational
principles. In texts, deviations from Zipf or Heaps laws reflect
semantic or syntactic organization.

A comparative science of component systems offers a unifying lens on different domains. Texts, for
instance, can be treated as assemblies of component words whose
statistical patterns—embodied in Heaps’ and Zipf’s laws— show the
balance between innovation and reuse as well as deeper semantic
dependencies. LEGO kits offer a complementary engineered setting in
which component diversity, design constraints, and creativity can be
quantified~\cite{mazzolini2018statistics}. In
biological contexts, microbiomes present functional gene repertoires
as components whose organization follows macroecological laws
observable in metagenomic data \cite{grilli2020macroecological}. Single-cell RNA-seq
provides a parallel cellular perspective, with genes acting as
components whose usage varies across cells
\cite{valle2020topic, lazzardi2023emergent} and  ultimately defines the cell identity~\cite{quake2021cell,biondo2025intrinsic}.
Together, these arenas enable
robust tests of the generality of statistical-mechanical approaches
and help reveal which features of component systems—growth,
dependency, innovation, or constraints—drive both their shared
signatures and their domain-specific differences.

A vast plethora of theoretical models have been proposed to explain the origin of one or more of the  statistical laws we discussed. From stochastic processes based on realization growth by duplication and  innovation
dynamics — including processes based on the  sample-space dynamics   or on correlated novelty
processes \cite{corominas2015understanding,tria2014dynamics, iacopini2020interacting,mazzolini2018heaps} - to more abstract statistical mechanics descriptions based on the presence of latent or unobserved variables ~\cite{schwab2014zipf,aitchison2016zipf,marsili2013sampling}.  

A clear understanding of the formal connections or the crucial differences between these alternative theoretical descriptions is still missing. Statistical properties of empirical component systems that go beyond the basic laws we described (such as fluctuation properties or  correlation patterns) can represent a new testing ground for  model selection.

\section*{Conclusions and perspectives}

Taken together, these results suggest a general strategy for the study of component systems. Robust large-scale regularities emerge from minimal assumptions about sampling, combinatorics, and growth, defining a baseline notion of universality across domains. Against this baseline, systematic deviations—whether driven by dependencies, functional constraints, or history-dependent innovation—become the primary carriers of mechanistic and causal information. This perspective shifts the focus from an observational catalog of empirical laws to a quantitative framework for comparing generative mechanisms, inferring hidden structure, and identifying system-specific trends. It also raises a forward-looking methodological question: how can such baselines and deviations be learned, represented, and interrogated in increasingly complex, high-dimensional systems?

To address this question, Large Language Models (LLMs) introduce an unprecedented additional perspective. Trained on massive corpora, LLMs internalize statistical patterns of component systems such as language, code, and biological sequences. They can be used both as tools to extract latent structures (e.g., embeddings reflecting dependency relations) and as empirical systems in their own right, allowing us to probe what statistical and generative principles they have learned. We envision an emerging synthesis where null models, mechanistic generative models, and
LLM-based representation learning jointly inform a unified theory of
component systems.

Notably, Zipf-like distributions may emerge in machine learning from the inference process in under-sampled, high-dimensional conditions~\cite{Marsili2022,marsili2013sampling}. We implicitly assumed here that in component systems similar distributions arise already at the level of the system itself, as a consequence of heterogeneous components and intrinsic modular structure. This may be very likely for some systems (e.g., LEGO sets) but it is less clear for others (e.g. texts). 
In other words, Zipf’s law may be a hybrid phenomenon: partly a reflection of the system’s intrinsic modular organization and partly a consequence of strong under-sampling and optimal inference or learning dynamics. In the latter case, the observation of a Zipf law should be interpreted as a signature of the relevance of the measured variables, rather than a direct indicator of any particular system behavior, but the joint marginal laws would still be observable in data. Disentangling these contributions for specific empirical systems will be informative on their nature.

Despite progress, significant challenges remain. A key theoretical gap
is a unified statistical mechanical theory that classifies the diverse
generative models for component systems, bridging equilibrium
sampling, history-dependent processes, and rule-based assembly. Both
theoretically and empirically we lack systematic methods to studying
the granularity (``coarse-graining'') of components across
scales. Equally, the implications of component-level laws for
larger-scale phenomena, such as for example the interplay between genome
composition and microbiome function or resilience, remain largely
uncharted territory.


To sum up, the central theme of this perspective is that~\textit{null models} based on sampling, diversification, and innovation provide a shared language to compare disparate component systems. At the same time,
systematic deviations from these null trends reveal the key processes
-biological, cognitive, technological, ecological etc.- that make each
system unique.
Component systems provide a unifying lens for studying complexity
across disciplines. By combining null models, inference tools and
insights from modern machine learning, we can build a coherent
statistical mechanics of component systems. Such a framework has the
potential to clarify the origins of observed regularities, reveal the
mechanisms that shape complex architectures, and guide data-driven
discovery in several scientific domains.

\section*{Acknowledgments}
This work was supported by the Italian "Ministero dell’Università e della Ricerca", PRIN 2022 – COD. 2022PY8MHN – GeCoS: Genomic Component Systems.

\bibliography{bibliography}
\bibliographystyle{unsrt}

\end{document}